\documentclass[]{spie} 

\usepackage{amsmath,amsfonts,amssymb}
\usepackage{graphicx}
\usepackage[colorlinks=true, allcolors=blue]{hyperref}
\usepackage{subcaption}
\usepackage{siunitx}
\usepackage{hhline}
\usepackage{array}
\usepackage{placeins}

\newcommand{\RNum}[1]{\uppercase\expandafter{\romannumeral #1\relax}}

\title{Spectroscopic performance of the electrical functional models for the eXTP SFA-T detectors}

\author[a]{A. Altmann}
\author[a]{R. Andritschke}
\author[a]{V. Antonelli}
\author[a]{T.F. Bechteler}
\author[a]{V. Burwitz}
\author[a]{D. Fink}
\author[c]{C. Fiorini}
\author[a]{M. Goretti}
\author[a]{P. Knopp}
\author[b]{P. Lechner}
\author[a]{N. Meidinger}
\author[a]{J.W. Muegge}
\author[b]{C. Sandow}
\author[a]{R. Strecker}

\affil[a]{Max-Planck-Institut für extraterrestrische Physik, Gießenbachstr. 1, 85748 Garching, Germany}
\affil[b]{Halbleiterlabor der Max-Planck-Gesellschaft, Isarauenweg 1, 85748 Garching, Germany}
\affil[c]{Politecnico di Milano, Piazza Leonardo da Vinci 32, 20133 Milano, Italy}

\authorinfo{Alexander Altman, E-mail: aaltmann@mpe.mpg.de, Telephone:  +49 (0) 89 30 000 3225}

\begin{document} 
\maketitle

\begin{abstract}
The Spectroscopy Focusing Array (SFA) is one of the three instruments on the eXTP satellite. It consists of six telescopes plus focal plane cameras, where five are equipped with a Silicon Drift Detector (SDD) array. These five SFA-T (T stands for timing) instruments are used for observations with high time resolution
($\leq$ 10 µs), high throughput (dead time $\leq$ 5 \% when observing a 1 Crab source), and good energy resolution ($\leq$ 180 eV at 6~keV).
This paper presents the spectroscopic performance of the Electrical Functional Modules (EFM) for SFA-T that are assembled and tested at the Max Planck Institute for Extraterrestrial Physics (MPE). A focus lies on the energy resolution at the planned sensor operating temperature of $\SI{-45}{\degreeCelsius}$.
Additionally, the use of a baffle that covers the pixel edges is studied.

\end{abstract}


\section{INTRODUCTION}
\label{sec:intro} 

The eXTP mission is a Chinese space mission with the goal of studying fundamental physics under extreme conditions of density, gravity, and magnetism \cite{Zhang2025}. It features three main instruments, the Polarimetry Focusing Array (PFA), the Spectroscopy Focusing Array (SFA), and a Wideband Camera (W2F). Both PFA and SFA use identical Wolter-I mirrors. 
The SFA consists of six individual telescopes. One, the SFA imaging (SFA-I), is equipped with a PN-CCD \cite{Meidinger}. The other five telescopes, which are part of SFA timing (SFA-T), are equipped with a Silicon drift detector (SDD) array, and are used for high time resolution spectroscopy. 
SDDs are well-suited for high count rate X-ray spectroscopy \cite{LECHNER2001281} and have already been used in space \cite{Gendreau2016}.
\newline
In this paper, we give an update on the Electrical Functional Model (EFM) for SFA-T. First, the design of the EFM detector module is shown, highlighting changes from the Breadboard model.
Next, we present an update to the operating mode and report the energy resolution of the EFM in comparison with the requirement of $\leq\SI{180}{\electronvolt}$ at a photon energy of $\SI{6}{\kilo\electronvolt}$. Additionally, the use of a baffle to shadow pixel edges is discussed. Such a baffle can avoid charge sharing events and therefore lower the background due to blocking photons that would hit the pixel edges.

\section{Detector Module Design}
\label{sec:desi_det}
\begin{figure} [htbp]
	\begin{center}
		\begin{tabular}{c} 
			\includegraphics[width=0.99\linewidth]{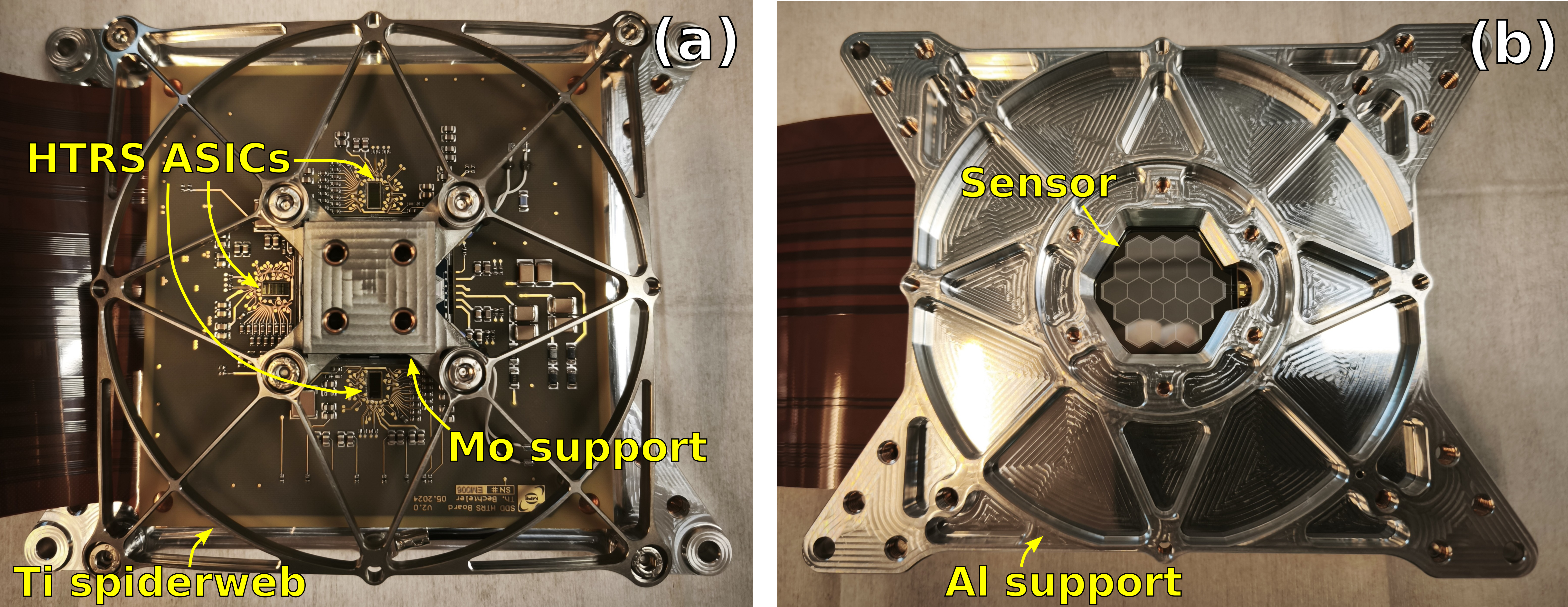}
		\end{tabular}
	\end{center}
	\caption[example] 	
	{\label{fig:dm}\textbf{(a)} Front side of the EFM detector module. The readout ASICs are located on the PCB, which is glued onto an Aluminum support structure.  \textbf{(b)} Back side of the EFM detector module. The SDD sensor is at the middle of the module. It has 19 hexagonal pixels with an area of $\SI{26.6}{\square\milli\meter}$ each (total sensitive area: $\SI{5.05}{\square\centi\meter}$)} 
\end{figure}

Going from Breadboard \cite{Altmann2024a} to Electrical Functional Model (EFM), several changes have been made in the mechanical and electrical design. The main driver for this was to get closer to a flight-like detector module design. Fig. \ref{fig:dm} shows the (a) front and (b) back sides of the EFM. 
The front side contains three High Time Resolution Spectroscopy Application-Specific Integrated Circuits (HTRS ASICs)\cite{Bombelli2013} used to read out signals from the sensor. The ASICs are glued onto a printed circuit board (PCB).  \newline
In addition to the applied ESA layout design rules (ECSS-Q-ST-70-12C \cite{ECSS_Q_ST_70_12C}), the routing on this PCB has been improved from the Breadboard version. The distance between signal lines has been increased to minimize coupling. In particular, care was taken to ensure that
the digital clock signals for the HTRS ASICs were separated as much as possible from the
analog signals coming from the SDD sensor. 
The PCB itself is glued onto an Aluminum support structure that provides the mounting points for the surrounding satellite structure.
The SDD sensor is glued to a support structure made of Molybdenum, which has a thermal expansion coefficient close to that of Silicon. 
A thin Titanium "spiderweb" structure connects the two support structures. 
This provides a high stiffness while maintaining a weak thermal coupling between sensor and PCB. The latter allows for cooling the sensor further down (e.g., \SI{-45}{\degreeCelsius}) compared to the PCB (e.g., \SI{20}{\degreeCelsius}). 
The mechanical design choices were influenced by knowledge gained from the SPICE project \cite{Muegge}.
\newline
The SDD sensor used for the detector modules has a sensitive area of $\SI{5.05}{\square\centi\meter}$, divided into 19 hexagonally shaped pixels with an area of $\SI{26.6}{\square\milli\meter}$ each. The sensor has a thickness of $\SI{450}{\micro\meter}$ and is fully depleted via the principle of sidewards depletion\cite{Gatti1984}. The photon entrance window, including a thin Aluminum layer to block light mainly in the visible energy range, is facing towards the viewer in Figure \ref{fig:dm}(b).

\section{Operating Mode}
\label{sec:opm}
\begin{figure} [htbp]
	\begin{center}
		\begin{tabular}{c} 
			\includegraphics[width=0.65\linewidth]{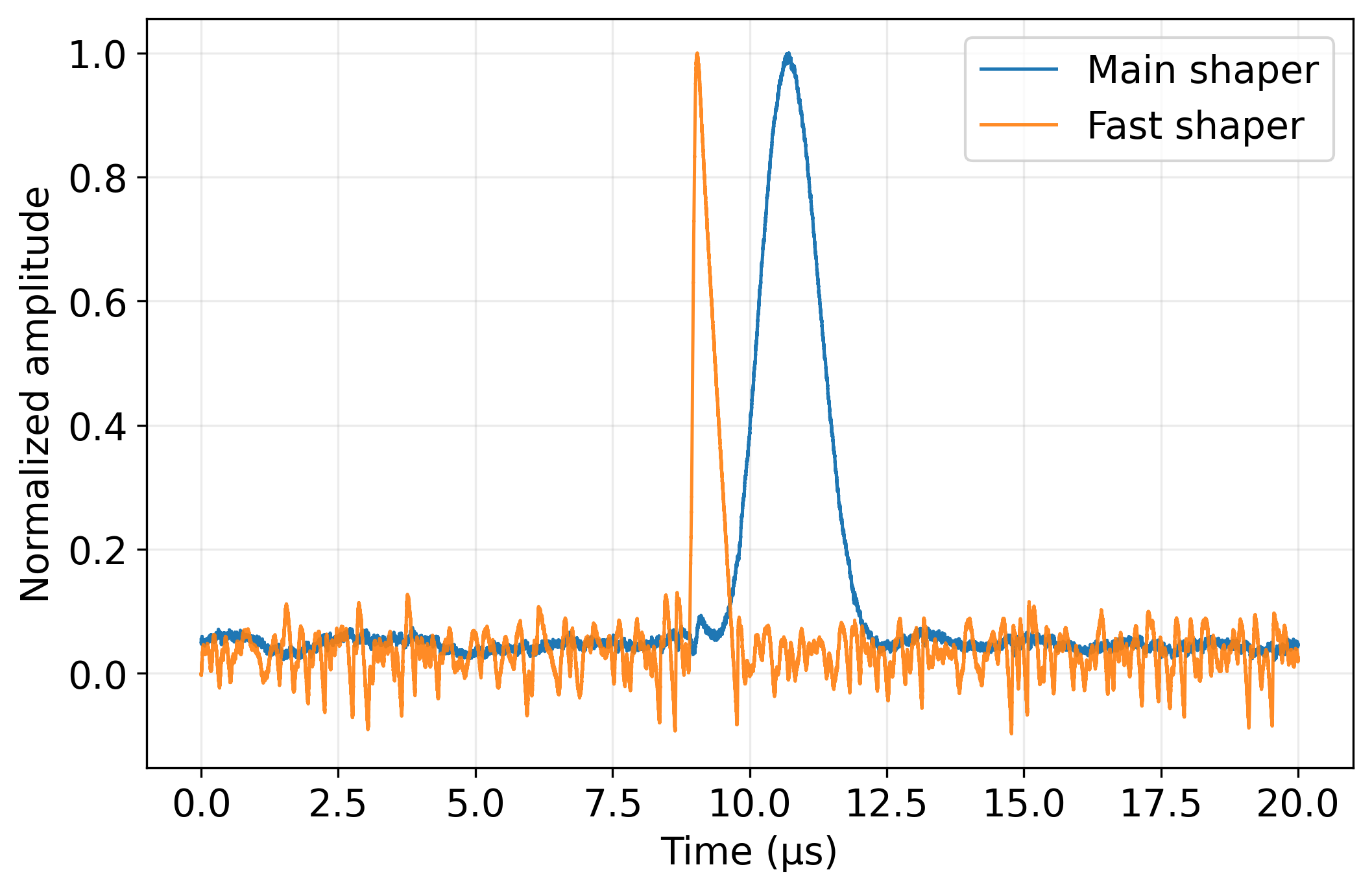}
		\end{tabular}
	\end{center}
	\caption[example] 	
	{\label{fig:osziscrs} Main shaper and fast shaper monitoring outputs of the HTRS ASIC recorded with an oscilloscope for one channel. At $\SI{8.5}{\micro\second}$, a photon event occurs. The 1 $\sigma$ noise determined from the baselines after normalization is  0.0090 a.u. for the main shaper and 0.0335 a.u. for the fast shaper.} 
\end{figure}

If a photon hits a pixel of the sensor, it generates electron-hole pairs in a number proportional to the photon energy. These get separated by the electric field inside the depleted silicon bulk. The electrons then drift towards the readout anode in the center of the pixel.
A charge-sensitive amplifier (CSA) readout structure is used, where the first field-effect transistor (FET) is located in the center of each pixel, and a preamplifier is located in each channel of the HTRS ASIC. Together with the feedback capacitance implemented on the sensor for every pixel, a CSA is formed, converting electrons to a voltage step at the output.
Inside the HTRS ASIC, this voltage step is then amplified and shaped into a semi-gaussian pulse. The amplitude of the pulse is proportional to the photon energy and is determined with an implemented peak-stretching circuit.
\newline 
The pile-up rejection mode of the HTRS ASIC has been described extensively in past works \cite{Bombelli2013}\cite{Altmann2024}. In principle, the ASIC discards events if they are closer in time than a programmable time setting. This can be used to avoid the recording of wrong photon energies in case of two photons arriving close in time. In that case, the shaped signals of both events in the ASIC disturb each other's amplitudes. Events are defined by the fast shaper signal crossing a set threshold. Then, a trigger signal is generated, and a time window, denoted as T2, starts. If another event is recognized before T2 has passed, the second signal gets discarded, and the time window T2 is re-triggered. A problem occurs when the fast shaper signal is too noisy. Then, noise excesses can overcome the threshold and trigger T2. If such a noise excess is in the vicinity of a real photon event, the photon event would get lost. The combination of a high fast shaper noise and a low threshold set in the ASIC would therefore lead to a significant loss of throughput. 
Since one of the goals of eXTP SFA is detecting low-energy photons down to 0.5 keV, the thresholds have to be set quite low.
Our investigations indicate that the noise in the fast shaper signal is too high to be able to do that. Fig. \ref{fig:osziscrs} shows the typical signal in the fast and main shaper of one channel of the ASIC. The shaping times are $\SI{1.4}{\micro\second}$ for the main shaper and $\SI{0.14}{\micro\second}$ for the fast shaper.
  \newline
For both shapers, debug outputs are available, and they can be investigated, e.g., with an oscilloscope. 
In the picture, a photon signal (at around $\SI{8.5}{\micro\second}$) is processed by the shapers, which leads to a significant semi-gaussian peak. The amplitudes were normalized to 1 for better comparison of both signals. In case of no event, one can easily see that the baseline of both shapers differs significantly. The 1 $\sigma$ noise of the baseline signal is 0.0335 a.u. for the fast shaper and 0.0090 a.u. for the main shaper. \newline
When the noise of the fast shaper limits the low energy range as described before, it is possible to switch off the pile-up rejection logic of the ASIC. 
Then, only the main shaper signal, which is significantly less noisy, is considered to determine events.
The disadvantage of this is, of course, that the pile-up rejection is not used to discard wrong photon energies. But this is mainly an issue at high count rates, while also depending on the shaping time. A longer shaping time increases the possibility of pile-up.
In Section \ref{sec:perf}, we compare two measurements at different count rates and show the difference in pile-up depending on the count rate.

\section{Spectrum and Energy Resolution}
\label{sec:perf} 

\begin{figure} [htbp]
	\begin{center}
		\begin{tabular}{c} 
			\includegraphics[width=0.99\linewidth]{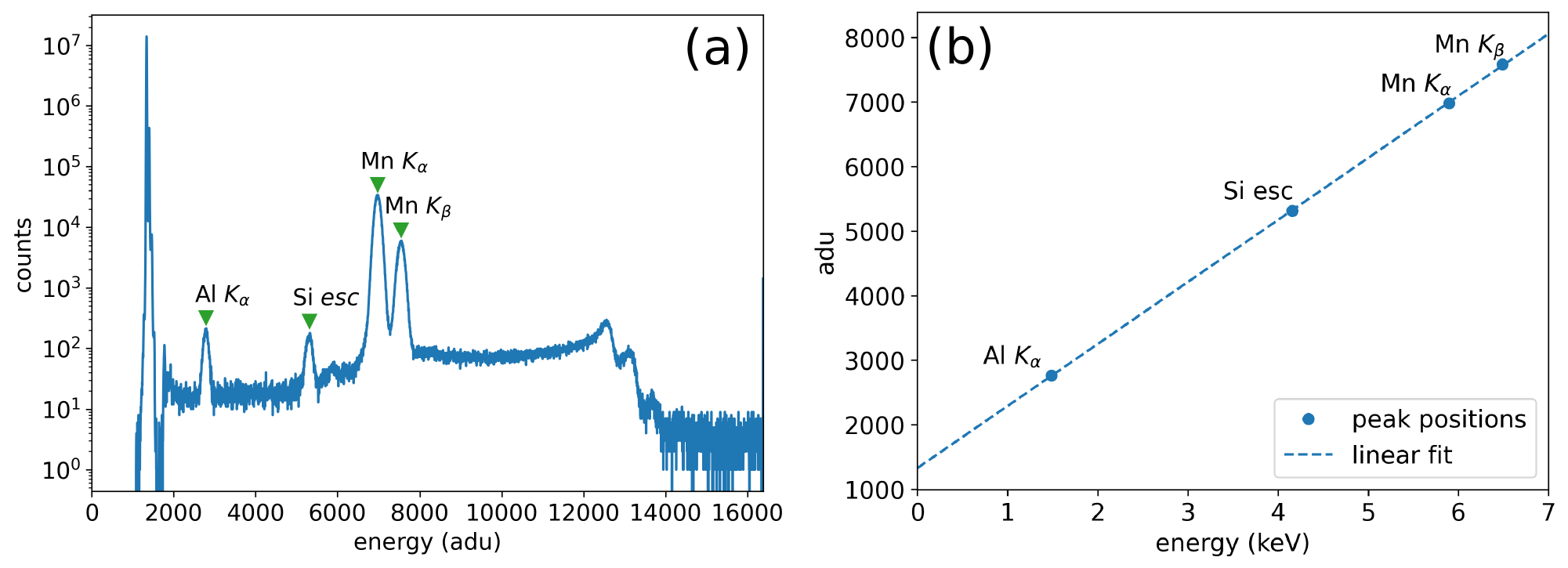}		
		\end{tabular}
	\end{center}
	\caption[example] 	
	{\label{fig:offset}Offset Determination of the central pixel of the EFM. \textbf{(a)} From the raw data spectrum, an algorithm determines significant peaks. The exact peak position is determined by a Gaussian fit. \textbf{(b)} The peak positions in ADU are plotted against the corresponding energies in keV. By applying a linear fit, the offset can be determined. It is equal to the value of the fit function at 0 eV.} 
\end{figure}

\begin{figure} [htbp]
	\begin{center}
		\begin{tabular}{c} 
			\includegraphics[width=0.8\linewidth]{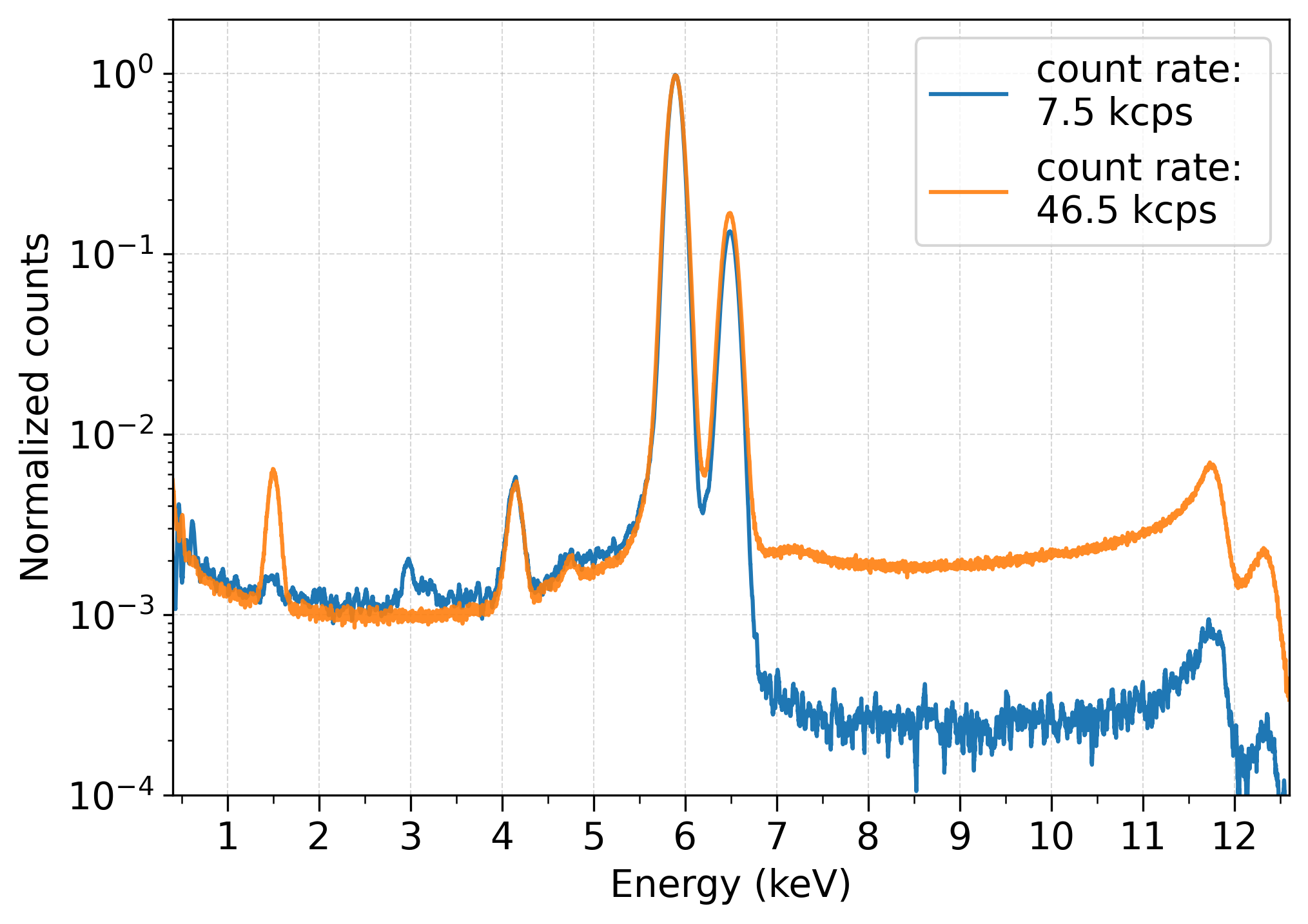}
		\end{tabular}
	\end{center}
	\caption[example] 	
	{\label{fig:specrum_all}Normalized spectrum of all 19 pixels combined, obtained with the EFM. Using a $^{55}$Fe source, the prominent peaks are Mn-K$_{\alpha}$, Mn-K$_{\beta}$, and the escape peaks of Mn-K$_{\alpha}$ and Mn-K$_{\beta}$. Because of an Al-Baffle and the use of Al foil in front of the X-ray source in case of the high count rate measurement, Al-K$_{\alpha}$ is also visible. The blue spectrum is measured with a count rate of 7.5 kcps recorded in the central pixel. The orange spectrum is measured with a count rate of 46.5 kcps recorded in the central pixel. At the higher count rate, pile-up is more likely, which leads to a higher background above the Mn-K$_{\beta}$ peak.}
\end{figure}

\begin{figure} [htbp]
	\begin{center}
		\begin{tabular}{c} 
			\includegraphics[width=0.8\linewidth]{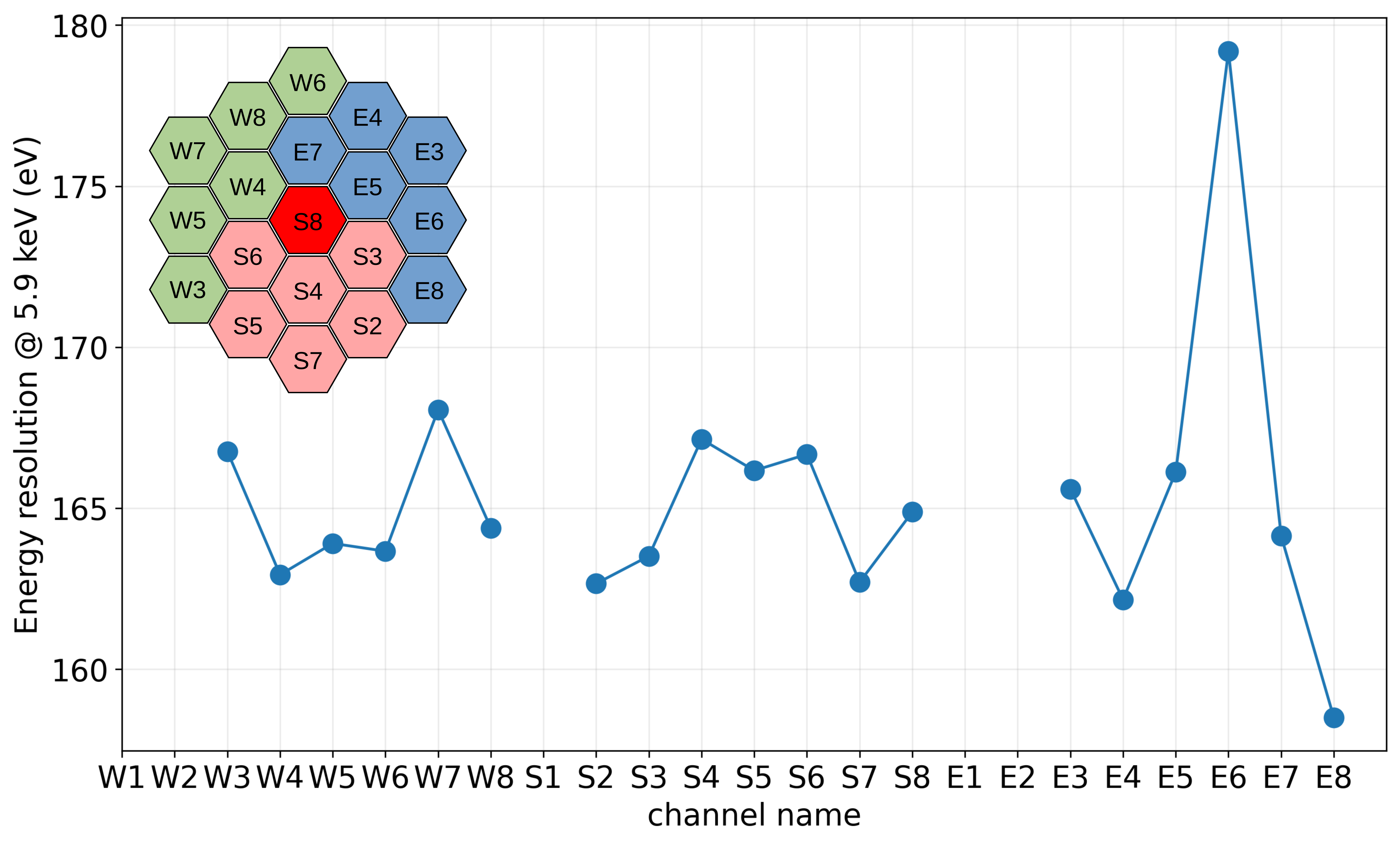}
		\end{tabular}
	\end{center}
	\caption[example] 
	{\label{fig:enres} Energy resolution of all pixels of the EFM. The inlay shows where each pixel is located on the sensor when looking from the front side onto it. The letters indicate which ASIC (West, South, East) and the numbers indicate which channel of the ASIC the pixel is connected to. The mean energy resolution of all pixels is $ \SI{165.2}{\electronvolt} $ at $ \SI{5.9}{\kilo\electronvolt} $. Temperature: $\SI{-45}{\degreeCelsius}$, shaping time: $\SI{1.4}{\micro\second}$, count rate: $\SI{7.5}{kcps}$ in the central pixel.} 
\end{figure}

When analyzing measured data, the electronic offset has to be determined first. A common method to determine the offset is to take dark frame measurements, which means measurements without a photon source emitting photons onto the detector. By averaging the signal in each channel over the whole measurement period, the offset can be determined and can then be subtracted from the measurement data with photons channel-wise.
If we use the same method with our detector, the offset is incorrect. This becomes clear when attempting to fit the Mn-K$_\alpha$ and Mn-K$_{\beta}$ peaks with a fit function that assumes the ratio of both peak positions to be 0.91. The fit diverges significantly from the measured spectra, indicating an incorrect offset.
\newline
To determine the correct offset, we have to analyze the raw data spectrum 
first. As an example, Figure \ref{fig:offset}(a) shows the raw data spectrum of the central pixel of the EFM.
From this spectrum, we can then identify peaks. 
In our case, using a $^{55}$Fe source, the prominent peaks are the Mn-K$_\alpha$ and Mn K$_\beta$ emission lines, and the Mn-K$_\alpha$ Si-escape peak. Because the $^{55}$Fe source in use is very strong, we put a few layers of Aluminum foil in front of the $^{55}$Fe source to reduce the photon flux on the detector. A positive side effect of the Aluminum foil is that it introduces Al-K$_{\alpha}$ fluorescence photons and therefore a corresponding peak in the obtained spectrum. This peak is useful for the offset determination.
As a first step, a peak finding algorithm identifies peaks in the raw data histograms of a pixel. Then, the exact position of the peaks, indicated by the green markers in Figure \ref{fig:offset}(a), is determined by a Gaussian fit.
Afterward, the peaks are ordered by height and assigned to the corresponding energies.
In the last step, we plot the ADU values of the found peaks against the energy in eV, where the peaks are located. With a linear fit, we can then determine the offset in ADU, which is the ADU value that corresponds to 0 eV (see Figure \ref{fig:offset}(b)).
This method works better the more peaks are available, especially when the peaks are far apart in terms of energy. Nonlinearities in the detector would make the offset determination more difficult, but our detector seems to be very linear in the observed energy range.
This whole process of determining the offset is implemented in our analysis software using the ROOT-based Offline Analysis (ROAn) framework \cite{Lauf2013}. \newline
After subtracting the correct offset, the gain of each pixel is determined. For this, the Mn-K$_\alpha$ and Mn-K$_\beta$ peaks are fitted with a Gaussian. The positional value of the Mn-K$_\alpha$ peak in units of ADU divided by the energy of Mn-K$_{\alpha}$, $\SI{5.9}{\kilo\electronvolt}$,
is the gain in that pixel. Combining all gain-corrected spectra of the 19 individual pixels yields a sum spectrum as presented in Fig. \ref{fig:specrum_all}.     
Most of the visible peaks have been described earlier, see also Fig. \ref{fig:offset}(a). 
The smaller double peak around 3 keV could be the Ag-L$_{\alpha}$ and Ag-L$_{\beta}$ lines.
Therefore, Silver has to be present in the surroundings of the radioactive source, e.g. in its housing. \newline
The plot shows measurements at two different count rates.
In the low count rate measurement, 7.5 kilocounts per second (kcps) were measured in the central pixel. For this measurement, an older, weaker X-ray source was used. In the high count rate measurement, 46.5 kcps were recorded in the central pixel. For this measurement, a new and stronger source was used, which was also attenuated with a few layers of Aluminum foil, for a total thickness of approximately $\SI{48}{\micro\meter}$. 
This explains why the Al peak is much more prominent here, since X-ray fluorescence photons are generated in the Aluminum foil. Also, since the absorption probability of photons in the foil decreases with increasing X-ray energy, the increase of the Mn-K$_{\beta}$ peak height relative to the Mn-K$_{\alpha}$ peak can be explained. The same explanation could apply to the absence of an Ag peak in the high count rate measurement.
But this could also be due to differences in the housing between the two radioactive sources. Beyond the Mn-K$_{\beta}$ peak, the background is significantly higher for the high count rate measurement, since there is an increased probability for pile-up. Small peaks appear at around $\SI{11.8}{\kilo\electronvolt}$ and $\SI{12.4}{\kilo\electronvolt}$, which correspond to the energy of two Mn-K$_{\alpha}$ or one Mn-K$_{\alpha}$ and one Mn-K$_{\beta}$ photon.  \FloatBarrier \noindent
Fig. \ref{fig:enres} shows the energy resolution as full-width half maximum (FWHM) of every pixel at $\SI{5.9}{keV}$ for the shaping time $\SI{1.4}{\micro\second}$ at $\SI{-45}{\degreeCelsius}$ and a count rate of 7.5 kcps in the central pixel after optimization of the bias voltages. 
The mean energy resolution equals $\SI{165.2}{\electronvolt}$. 
%On a satellite, cooling power is limited due to limited radiator area. 
Increasing the sensor temperature to, e.g., $\SI{-30}{\degreeCelsius}$ does not significantly worsen the energy resolution. At $\SI{-30}{\degreeCelsius}$, the mean energy resolution is $\SI{165.4}{\electronvolt}$. But we have to account for potential radiation damage during the mission lifetime. 
Such damage can lead to an increased dark current, which could then be compensated by a lower temperature. Thus, the target temperature for the SDD sensor in operation is $\SI{-45}{\degreeCelsius}$.
All pixels satisfy the requirement of less than 180 eV at 6 keV. One pixel, E6,  performs significantly worse. It is not understood why, although smaller variations between individual pixels are expected.
For the first sensor we tested, which was integrated on the breadboard module, it was the central pixel that performed the worst. It is good to know that this is not a systematic issue. The central pixel is the most important one since the majority of X-rays will be focused onto this pixel.
\FloatBarrier

\section{Baffle and back frame}
\begin{figure} [htbp]
	\begin{center}
		\begin{tabular}{c} 	
			\includegraphics[width=0.8\linewidth]{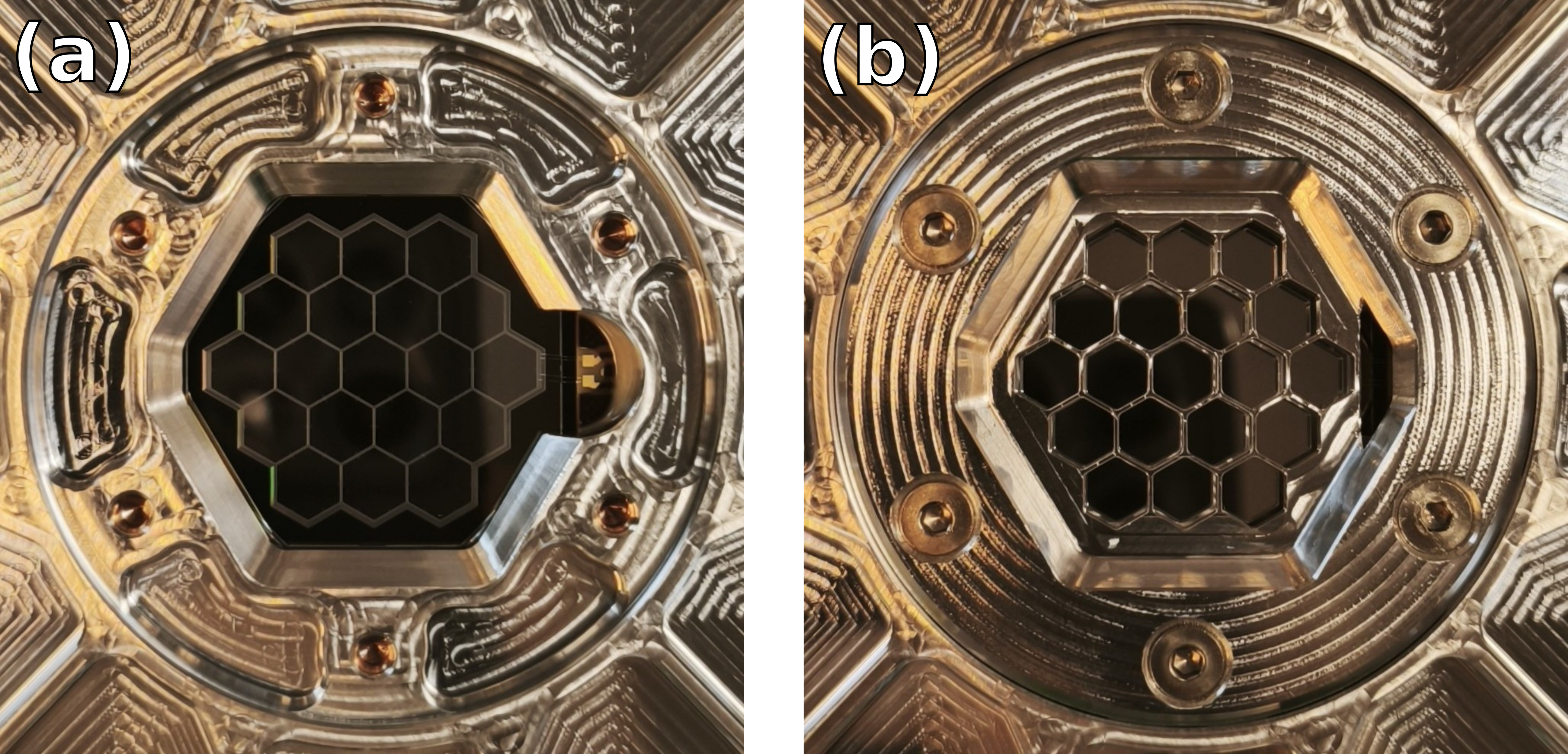}		
		\end{tabular}
	\end{center}
	\caption[example] 	
	{\label{fig:baffle}\textbf{(a)} Sensor without a baffle. \textbf{(b)} Sensor with a baffle made of Aluminum covering the pixel edges and therefore avoiding split events.} 
\end{figure}

\begin{figure} [ht]
	\begin{center}
		\begin{tabular}{c} 
			\includegraphics[width=0.8\linewidth]{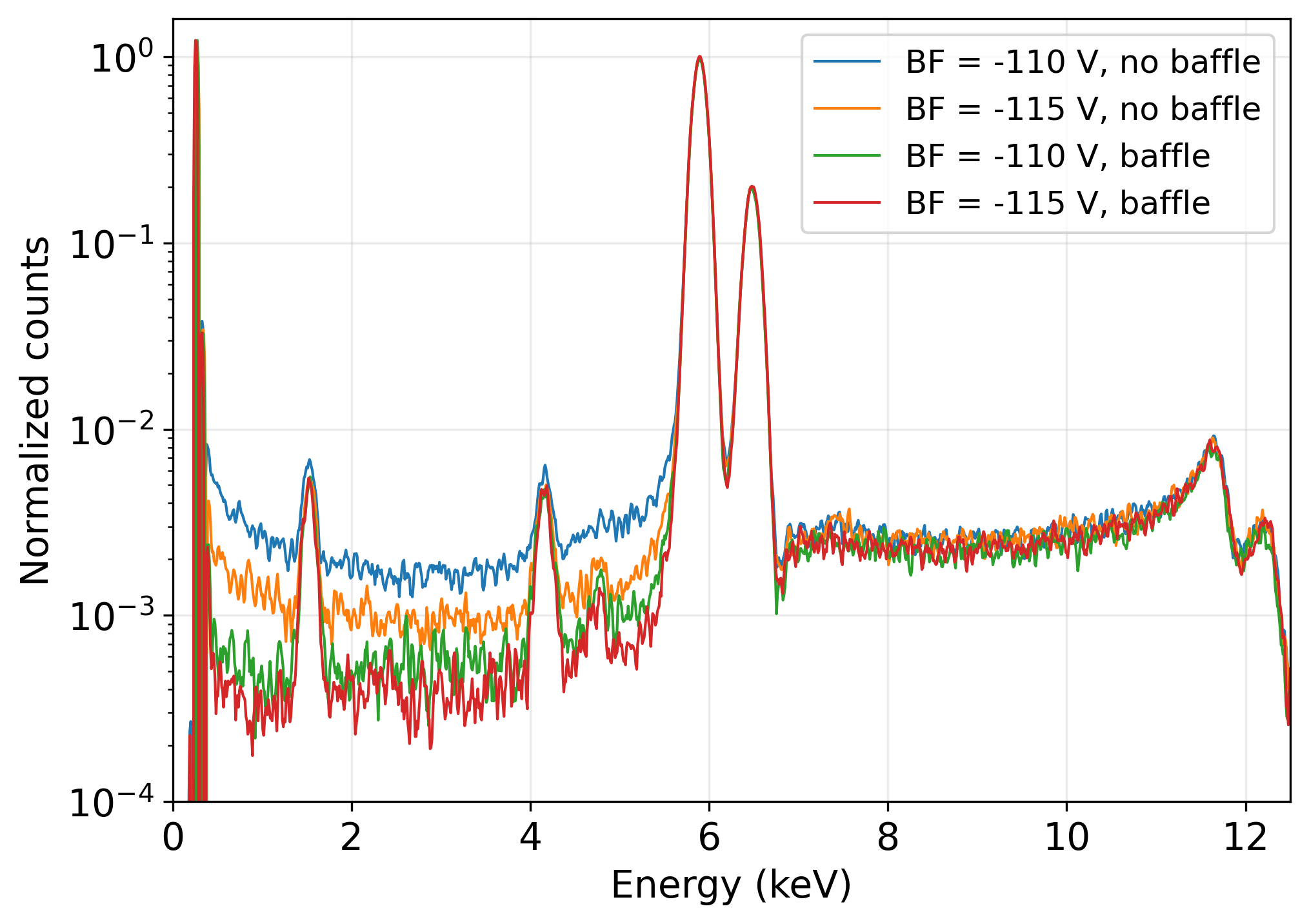}
		\end{tabular}
	\end{center}
	\caption[example] 	
	{\label{fig:baffle_on_off} Comparison of measured spectra in the central pixel for different back frame (BF) voltages. The voltage of the back contact was kept constant at $\SI{-110}{\volt}$. If the back frame voltage is more negative, the background is lowered. The use of a baffle that covers pixel edges has a similar effect by preventing split events.} 
\end{figure}

In a pixelated detector, charge sharing between adjacent pixels, leading to so-called split events, can occur when a photon hits the sensor at or close to the border between pixels. 
It is an option to recombine these split events. 
%(Tricky high count rates?)
But to avoid split events from occurring, a baffle can be used that covers pixel edges. Fig. \ref{fig:baffle} shows the center of the EFM from the entrance window side with the sensor in the middle, and (a) no baffle installed compared to (b) a baffle installed on the module.
 \newline
The material of this baffle is important because it affects how deep the baffle struts have to be made to have a sufficient stopping power for X-rays. Also, the material determines which fluorescence photons will appear in the measured spectrum when higher-energy photons hit the baffle.
The fluorescence photons should not be at energies that are important for astronomy.
The baffle version used here has thicker struts than necessary. This is due to the conditions in the current measurement chamber. The distance between the radioactive source and the detector is relatively small (around $\SI{8}{\centi\meter}$). This means the X-ray beam is divergent and could therefore bypass the baffle, especially for pixels that are located in the outer ring of the sensor.
As proof of concept, our baffle is sufficient. If it is planned to use a baffle with the flight modules, the dimensions have to be updated, because unnecessarily thick baffle struts would lead to a higher loss of photons. \newline
We took measurements with and without such a baffle and present the corresponding spectra in Fig. \ref{fig:baffle_on_off}. 
Due to divergent X-rays, especially at outer pixels, only the central pixel is shown. Taking into account  the position of the $^{55}$Fe source in relation to the sensor, the X-ray beam is expected to be the most parallel at the central pixel. This ensures the effect of the baffle is the strongest here.
One can see the positive effect of using a baffle, which becomes clear when we compare the background. With a baffle, it is noticeably reduced due to the reduction of split events.
 \newline
Another measure to lower the background is the use of the built-in back frame. Many Silicon sensors have a homogeneous entrance window with a gapless p+ implantation for the back contact \cite{Lechner1996}. The SDDs for eXTP SFA-T, however, also have a back frame. This is an additional p+ implantation on the back side of the chip that goes along the pixel edges, separated by a small gap from the back contact.
The purpose of the back frame is the following.
At the pixel edge close to the entrance window, the electric field is weak. If photons interact here, the generated electrons could then also diffuse to a neighboring pixel. If the voltage applied to the back frame is more negative than that to the back contact, the electric field at the pixel edge is stronger. Then, charge sharing between neigbouring pixels becomes less likely.
As shown in Fig. \ref{fig:baffle_on_off}, the background is therefore reduced with a more negative back frame (BF) voltage compared to the back contact voltage. The latter has been kept constant at $V_{BC} = \SI{-110}{\volt}$ while the back frame has been varied between $V_{BF} = \SI{-110}{\volt}$ and $V_{BF} = \SI{-115}{\volt}$. Making the back frame voltage more negative ($V_{BF} = \SI{-120}{\volt}$) did not provide a further improvement of the background. 
\newline

\section{summary and outlook}
In this paper, we presented the EFMs for SFA-T. The energy resolution of every pixel is within the requirement of better than 180 eV at 6 keV at a shaping time of $\SI{1.4}{\micro\second}$ and a temperature of $\SI{-45}{\degreeCelsius}$. We recommend using a baffle to avoid split events and lower the background.
The redesign phase for the qualification model has started, including changes to the mechanical and electrical design. \newline
Additionally, the design for the SDD flight production has been finalized. It includes minor changes to the layout.
For example, the readout structure will be made smaller, leading to a smaller anode area and therefore lower capacitance. This is expected to have a positive effect on the performance by providing a better signal-to-noise ratio. Production of the flight sensors will start in summer 2026.

\acknowledgments     

Development and production of the SDD sensors for eXTP SFA are performed in collaboration between MPE and the MPG Semiconductor Laboratory (HLL).
The authors want to thank Franz Soller and Patricia Langer for their help in assembling the EFM detector modules.
%The authors want to thank Patricia Langer and Franz Soller for preparing the SDD detector module presented here.
This work was funded by the Max Planck Society.

\bibliography{bibliography_spie_paper}
\bibliographystyle{spiebib}

\end{document}